\documentclass[twocolumn,showpacs,aps,pre]{revtex4}
\usepackage{amsmath}
\usepackage{amssymb}
\usepackage{graphicx}
\usepackage{graphics}

\input{tcilatex}

\begin{document}

\title{Bimodal distribution function of a 3d wormlike chain with a fixed
orientation of one end}
\author{F. F. Semeriyanov and S. Stepanow}
\affiliation{Institut f\"{u}r Physik, Martin-Luther-Universit\"{a}t Halle-Wittenberg,
D-06099 Halle, Germany}
\date{\today}

\begin{abstract}
We study the distribution function of the three dimensional wormlike chain
with a fixed orientation of one chain end using the exact representation of
the distribution function in terms of the Green's function of the quantum
rigid rotator in a homogeneous external field. The transverse 1d
distribution function of the free chain end displays a bimodal shape in the
intermediate range of the chain lengths ($1.3L_{p},...,3.5L_{p}$). We
present also analytical results for short and long chains, which are in
complete agreement with the results of previous studies obtained using
different methods.
\end{abstract}

\maketitle



\section{Introduction}

\label{intro}

Polymers with contour length $L$ much larger than the persistence length $%
L_{p}$, which is the correlation length for the tangent-tangent correlation
function along the polymer and is a quantitative measure of the polymer
stiffness, are flexible and are described by using the tools of quantum
mechanics and quantum field theory \cite{edwards65}-\cite{schaefer-book}. If
the chain length decreases, the chain stiffness becomes an important factor.
Many polymer molecules have internal stiffness and cannot be modeled by the
model of flexible polymers developed by Edwards \cite{edwards65}. The
standard coarse-graining model of a wormlike polymer was proposed by Kratky
and Porod \cite{kratky-porod49}. The essential ingredients of this model are
the penalty for the bending energy and the local inextensibility. The latter
makes the treatment of the model much more difficult. There have been a
substantial number of studies of the Kratky-Porod model in the last half
century \cite{hermans52}-\cite{yamakawa97} (and citations therein). In
recent years there has been increasing interest in the theoretical
description of semiflexible polymers \cite{wilhelm96}-\cite{stepanow05}. The
reason for this interest is due to potential applications in biology \cite%
{allemand05} (and citations therein) and in research on semicrystalline
polymers \cite{doi88}.

It was found in the recent numerical work\ by Lattanzi et al. \cite%
{lattanzi04}, and studied analytically in \cite{benetatos05} within the
effective medium approach, that the transverse distribution function of a
polymer embedded in two-dimensional space possesses a bimodal shape for
short polymers, which is considered to be a manifestation of the
semiflexibility. The bimodal shape for the related distribution function of
the 2d polymer was also found in recent exact calculations by Spakowitz and
Wang \cite{spakowitz05}. In this paper we study the transverse distribution
function $G(\mathbf{t}_{0},x=0,y,N)$ of the three dimensional wormlike chain
with a fixed orientation $\mathbf{t}_{0}$ of one polymer end using the exact
representation of the distribution function in terms of the matrix element
of the Green's function of the quantum rigid rotator in a homogeneous
external field \cite{stepanow04}. The exact solution of the Green's function
made it possible to compute the quantities such as the structure factor, the
end-to-end distribution function, etc. practically exact in the definite
range of parameters \cite{spakowitz04}, \cite{stepanow04}. Our practically
exact calculations of the transverse distribution function of the 3d
wormlike chain demonstrate that it possesses the bimodal shape in the
intermediate range of the chain lengths ($1.3L_{p},...,3.5L_{p}$). In
addition, we present analytical results for short and long wormlike chain
based on the exact formula (\ref{Gtkp}), which are in complete agreement
with the previous results obtained in different ways \cite{yamakawa73} (WKB
method for short polymer), \cite{gobush72} (perturbation theory for large
chain).

The paper is organized as follows. Section \ref{sect1} introduces
to the formalism and to analytical considerations for short and
large polymers. Section \ref{numer} contains results of the
numerical computation of the distribution function for polymers
with different number of monomers.

\section{Analytical treatment}

\label{sect1}

The Fourier-Laplace transform of the distribution function of the free end
of the wormlike chain with a fixed orientation $\mathbf{t}_{0}=$ $\left. d%
\mathbf{r}(s)/ds\right\vert _{s=0}$ of the second end is expressed,
according to \cite{stepanow04}, in a compact form through the matrix
elements of the Green's function of the quantum rigid rotator in a
homogeneous external field $\tilde{P}^{s}(k,p)$ as
\begin{equation}
G(\mathbf{t}_{0},\mathbf{k},p)=\underset{l=0}{\overset{\infty }{\sum }}%
\langle 0|\tilde{P}^{s}(k,p)|l\rangle \sqrt{2l+1}P_{l}(\mathbf{t}_{0}\mathbf{%
n}),  \label{Gtkp}
\end{equation}%
where $\mathbf{n}=\mathbf{k}/k$, and $\tilde{P}^{s}(k,p)$ is defined by
\begin{equation}
\tilde{P}^{s}(k,p)=\left( 1+ikDM^{s}\right) ^{-1}D,  \label{Ps}
\end{equation}%
with $D$ and $M^{s}$ being the infinite order square matrices given by
\begin{equation}
D_{l,l^{\prime }}=\frac{\delta _{l,l^{\prime }}}{p+\frac{l(l+1)}{2}},\ \
M_{l,l^{\prime }}^{s}=w_{l}\delta _{l,l^{\prime }+1}+w_{l+1}\delta
_{l+1,l^{\prime }},  \label{D}
\end{equation}%
and $w_{l}=\sqrt{l^{2}/(4l^{2}-1)}$. The matrix $D$ is related to the energy
eigenvalues of the free rigid rotator, while $M^{s}$ gives the matrix
elements of the homogeneous external field. Since $\tilde{P}^{s}(k,p)$ is
the infinite order matrix, a truncation is necessary in the performing
calculations. The truncation of the infinite order matrix of the Green's
function by the $n\times n$-order matrix contains all moments of the
end-to-end chain distance, and describes the first $2n-2$ moments exactly.
The transverse distribution function we consider, $G(\mathbf{t}_{0},x=0,y,N)$%
, is obtained from $G(\mathbf{t}_{0},x,y,z,N)$, which is determined by Eqs. (%
\ref{Gtkp})-(\ref{D}), integrating it over the $z$-coordinate, and imposing
the condition that the free end of the chain stays in the $x=0$ plane. As a
result we obtain
\begin{eqnarray}
&&G(\mathbf{t}_{0},x=0,y,N)=\underset{-\infty }{\overset{\infty }{\int }}dzG(%
\mathbf{t}_{0},x=0,y,z,N)  \notag \\
&=&\underset{0}{\overset{\infty }{\int }}kdk\,G(\mathbf{t}_{0},\mathbf{k}%
|_{k_{z}=0},N)J_{0}(ky),  \label{GyN}
\end{eqnarray}%
where $J_{0}(ky)$ is the Bessel function of the first kind \cite%
{AbramowitzStegun}. Taking the $z$-axis to be in the direction of $\mathbf{t}%
_{0}$ yields $\mathbf{t}_{0}\mathbf{n}=0$, so that the arguments of the
Legendre polynomials in Eq. (\ref{Gtkp}) become zero, and consequently only
even $l$ will contribute to the distribution function (\ref{GyN}).

\subsection{Short wormlike chain}

We now will consider the expansion of (\ref{Gtkp}) around the rod limit $%
N\rightarrow 0$, which corresponds to the expansion of $G(\mathbf{t},\mathbf{%
k},p)$ in inverse powers of $p$. To derive such an expansion, we write $D$
in the equivalent form as%
\begin{equation}
D=D^{0}-\frac{1}{p}ED  \label{Dit}
\end{equation}%
with $D_{l,l^{\prime }}^{0}=p^{-1}\delta _{l,l^{\prime }}$ and $%
E_{l,l^{\prime }}=l(l+1)/2\delta _{l,l^{\prime }}$. Further we introduce the
notation $\left( 1+ikDM^{s}\right) ^{-1}=y_{0}\tilde{y}$ with $y_{0}$ and $%
\tilde{y}$ defined by%
\begin{equation}
y_{0}=\left( 1+ikD^{0}M^{s}\right) ^{-1},\ \ \tilde{y}=1+ik\frac{E}{p}%
DM^{s}y_{0}\tilde{y}.\ \   \label{y}
\end{equation}%
The iteration of $D$ and $\tilde{y}$ results in the desired expansion of $%
\tilde{y}$ and consequently of $G(\mathbf{t}_{0},\mathbf{k},p)$ in inverse
powers of $p$, which corresponds to an expansion of $G(\mathbf{t}_{0},%
\mathbf{k},N)$ in powers of $N$. The leading order term in the short chain
expansion is obtained by replacing $D$ by $D^{0}$ in Eq. (\ref{Gtkp}) as
\begin{equation}
G_{0}(\mathbf{t}_{0},\mathbf{k},p)=\underset{l=0}{\overset{\infty }{\sum }}%
\left[ y_{0}D^{0}\right] _{0l}\sqrt{2l+1}P_{l}(\mathbf{t}_{0}\mathbf{n}).
\label{Gtkp0}
\end{equation}%
The latter coincides with the expansion of the plane wave \cite%
{landau-lifshitz3}%
\begin{equation}
e^{-ikr\cos \theta }=\sum_{l=0}^{\infty }i^{l}(2l+1)\left( \frac{r}{k}%
\right) ^{l}\left( \frac{1}{r}\frac{d}{dr}\right) ^{l}\frac{\sin kr}{kr}%
P_{l}(\cos \theta ),  \label{PlW}
\end{equation}%
where $\cos \theta =\mathbf{t}_{0}\,\mathbf{n}$ is the angle between the
tangent $\mathbf{t}_{0}$ and the wave vector $\mathbf{k}$. The connection of
$G_{0}(\mathbf{t}_{0},\mathbf{k},p)$ with the plane wave expansion is due to
the fact that the Kratky-Porod chain becomes a stiff rod in the limit of
small $N$. We have checked the equivalency between the plane wave expansion (%
\ref{PlW}) and the distribution function (\ref{Gtkp0}) term by term
expanding (\ref{Gtkp0}) in series in powers of the wave vector $k$. The arc
length $N$ is equivalent for stiff rod in units under consideration to the
chain end-to-end distance $r$. In the $r$ space the plane wave (\ref{PlW})
corresponds to the stiff rod distribution function
\begin{equation}
G(\mathbf{t}_0,r,N)|_{N\rightarrow 0}=\delta (x)\delta (y)\delta (z-N).
\end{equation}

The iteration of $D$ in (\ref{Dit}) and $\tilde{y}$ in (\ref{y}) generates
an expansion of $G(\mathbf{t}_{0},\mathbf{k},p)$ in inverse powers of $p$.
The corrections to the plane wave to order $1/p^{3}$ are obtained as%
\begin{eqnarray*}
&&G_{1}(\mathbf{t}_{0},\mathbf{k},p)=\underset{l=0}{\overset{\infty }{\sum }}%
\left[ -y_{0}\frac{E}{p}D^{0}+y_{0}\left( \frac{E}{p}\right) ^{2}D^{0}\right.
\\
&+&\left. {i}ky_{0}\frac{E}{p}D^{0}M^{s}y_{0}D^{0}\right] _{0l}\sqrt{2l+1}%
P_{l}(\mathbf{t}_{0}\mathbf{n})+....
\end{eqnarray*}

The above procedure yields for $l=0$ the short chain expansion of the
distribution function of the free Kratky-Porod chain, which was studied
recently in \cite{stepanow05}. Unfortunately, we did not succeed yet in
analytical evaluation of $G_{1}(\mathbf{t}_{0},\mathbf{k},N)$. Such
computation would be an interesting alternative to the treatment of the
short limit of the wormlike chain by Yamakawa and Fujii \cite{yamakawa73}
within the WKB method. Nevertheless, following the consideration in \cite%
{stepanow05} we succeeded in computing the anisotropic moments $\left\langle
\left( \mathbf{rt}_{0}\right) ^{n}\right\rangle $ for small $N$%
\begin{equation}
\left\langle \left( \mathbf{rt}_{0}\right) ^{n}\right\rangle =N^{n}\left( 1-%
\frac{n}{2}N+\frac{n(5n-1)}{24}N^{2}+...\right) .  \label{Rt_0}
\end{equation}%
The first-order correction coincide with that obtained in \cite{yamakawa73}
using the WKB method, while the second-order correction is to our knowledge
new. The higher-order terms in (\ref{Rt_0}) can be established in a
straightforward way using the present method. Note that the computation of $%
\left\langle \left( \mathbf{rt}_{0}\right) ^{n}\right\rangle $ does not
require the knowledge of the full distribution function $G(\mathbf{t}%
_{0},r,N)$.

\subsection{Large chain}

In studying the end-to-end distribution function for large $N$ we utilize
the following procedure. We expand first the expression $\langle 0|\tilde{P}%
^{s}(k,p)|l\rangle $ in powers of $k^{2}$. The structure of this expansion
for $l=0$ is presented in Table \ref{Table1}.
\begin{table}[tbp]
\caption{{}}
\label{Table1}%
\begin{ruledtabular}
\textbf{%
\begin{tabular}{lllll}
$ q: $ & \multicolumn{1}{c}{1} &
\multicolumn{1}{c}{2} & \multicolumn{1}{c}{3} &  \\
\hline
$k^{0}:$ & $\frac{1}{p}$ &  &  &  \\
$k^{2}:$ & $\frac{-w_{1}^2}{p^{2}}$ & $\frac{w_{1}^2}{p}$ & $-w_{1}^2$ &  \\
$k^{4}:$ & $\frac{w_{1}^4}{p^{3}}$ &
$\frac{-6w_{1}^4+w_{1}^2w_{2}^2}{3p^{2}}$ &
$\frac{27w_{1}^4-7w_{1}^2w_{2}^2}{9p}$ &  \\
$k^{6}:$ & $\frac{-w_{1}^6}{p^{4}}$ &
$\frac{9w_{1}^6-2w_{1}^4w_{2}^2}{3p^{3}}$ &
$\frac{-324w_{1}^6+120w_{1}^4w_{2}^2-w_{2}^2w_{1}^2w_{3}^2-6w_{2}^4w_{1}^2}{
54p^{2}}$ &  \\
$k^{8}:$ & $\frac{w_{1}^8}{p^{5}}$ &
$\frac{w_{1}^6w_{2}^2-4w_{1}^8}{p^{4}}$ &
$\frac{-117w_{1}^6w_{2}^2+w_{1}^4w_{2}^2w_{3}^2+270w_{1}^8+9w_{1}^4w_{2}^4}{
27p^{3}}$ &  \\
... &  &  &  & \\
\end{tabular}%
}
\end{ruledtabular}
\end{table}
The subseries in powers of $k^{2}$ in the $q${\small th} column are denoted
by $T_{{q}}^{l}$. Thus we have
\begin{equation*}
G_{l}(k,p)\equiv \sqrt{2l+1}\langle 0|\tilde{P}^{s}(k,p)|l\rangle
=\sum_{q}T_{{q}}^{l}.
\end{equation*}%
The series $T_{{q}}^{l}$ with small values $l$ and $q$ possess a simple
structure and can be summed up. For example $T_{0}^{0}$ and $T_{1}^{0}$ are
given by
\begin{equation*}
T_{1}^{0}=3\sum_{m=1}^{\infty }\frac{\left( -1\right) ^{m-1}w{_{{1}}}%
^{2\,m}k^{2\,m-2}}{p^{m}}\,=\frac{1}{k^{2}/3+p},
\end{equation*}%
\begin{eqnarray*}
T_{2}^{0} &=&\sum_{m=1}^{\infty }\frac{\left( -1\right) ^{m-1}w{_{{1}}}%
^{2\,m}k^{2\,m}}{p^{m}}m\left( 1+{\frac{w_{2}^{2}k^{2}}{p}}\right) \\
&=&{\frac{k^{2}\left( 45\,p+4\,k^{2}\right) }{15\left( k^{2}+3\,p\right) ^{2}%
}}.
\end{eqnarray*}%
While $T_{1}^{0}$ corresponds to the distribution function of the Gaussian
chain, $T_{q}^{0}$ give the $q${\small th} correction to the Gaussian
distribution. The inspection of the series for $T_{1}^{1}$ and $T_{2}^{1}$
shows that they are expressed by $T_{1}^{0}$ and $T_{2}^{0}$ as%
\begin{eqnarray*}
T_{1}^{1} &=&-i\,k\,T_{1}^{0}, \\
T_{2}^{1} &=&-\frac{3\,i\,p}{k}T_{2}^{0}.
\end{eqnarray*}%
However, it seems that there is no general recursion relation for $T_{{q}%
}^{l}$. The results of computations of $G_{l}(k,p)$ for $l=1,2,3,4$ by
taking into account $q=1,2,3$ ($l=0$) and $q=1,2$ ($l\neq 0$) are summarized
in Table \ref{Table2}.
\begin{table}[tbp]
\caption{{}}
\label{Table2}%
\begin{ruledtabular}
\begin{tabular}{cl}
$l$ & \multicolumn{1}{c}{$G_{l}(k,p)$} \\ \hline
0 : & $\frac{1}{k^{2}/3+p}+{\frac{{k}^{2}\left( 45\,p+4\,{k}^{2}\right) }{%
15\left( {k}^{2}+3\,p\right) ^{2}}} -{\frac{{k}^{2}\left( 4725\,{p}^{3}+90\,{%
k}^{4}p+980\,{k}^{2}{p}^{2}+2\,{k}^{6}\right) }{525\left( {k}%
^{2}+3\,p\right) ^{3}}}...$ \\
1 : & ${\frac {-3\,ik}{{k}^{2}+3\,p}} - {\frac {1/5\,ikp \left( 4\,{k}%
^{2}+45\,p \right) }{ \left( {k}^{2}+3\, p \right) ^{2}}}...$ \\
2 : & ${\frac {{k}^{2}}{3{k}^{2}+9\,p}} + {\frac {1}{210}}\,{\frac
{{k}^{2} \left( {k}^{4}+45\,{k}^{2}p+280\,{p } ^{2} \right) }{
\left( {k}^{2}+3\,p
\right) ^{2}}}...$ \\
3 : & ${\frac {i{k}^{3}}{15\,{k}^{2}+45\,p}}+{\frac {1}{2025}}\,{\frac {i{k}%
^{5} \left( 65\,{k}^{2}+294\,p \right) }{ \left( {k}^{2}+3\,p \right) ^{2}}}%
...$ \\
4 : & ${\frac {2}{525}}\,{\frac {{k}^{4}}{{k}^{2}+3\,p}}+{\frac {1}{259875}}\,{%
\frac {508\,{k}^{8}+2250\,{k}^{6}p}{ \left( {k}^ {2}+3\,p \right)
^{2}}}...$
\end{tabular}
\end{ruledtabular}
\end{table}

The inverse Laplace-Fourier transform of $G_{l}(k,p)$ given in the table
yields the expansion of the end-to-end distribution function $G(R,\Theta ,t)$
to order $O(t^{-2})$ as
\begin{eqnarray}
&&G(R,\Theta ,t)=(4\pi )^{-1}\left( \frac{3}{2\,\pi \,t}\right) ^{3/2}\exp
\left( -\frac{3\,R^{2}}{2\,t}\right) \times  \notag \\
&&\left[ 1-\frac{5}{8\,t}+2\,{\frac{R^{2}}{{t}^{2}}}-{\frac{33}{40}}\,{\frac{%
R^{4}}{{t}^{3}}}+{\frac{6799}{1600}}\,{\frac{R^{4}}{{t}^{4}}}\right.  \notag
\\
&&-{\frac{3441}{1400}}\,{\frac{R^{6}}{{t}^{5}}}+{\frac{1089}{3200}}\,{\frac{%
R^{8}}{{t}^{6}}}-{\frac{329}{240}}\,{\frac{R^{2}}{{t}^{3}}}-{\frac{79}{%
640\,t^{2}}}  \notag \\
&&+(\frac{3}{2}{\frac{R}{t}}+{\frac{153}{40}}\,{\frac{R^{3}}{{t}^{3}}}-{\
\frac{99}{80}}\,{\frac{R^{5}}{{t}^{4}}}-{\frac{25}{16}}\,{\frac{R}{{t}^{2}}}%
)P_{1}(\cos \Theta )  \notag \\
&&+(\frac{1}{2}{\frac{R^{2}}{{t}^{2}}}+{\frac{961}{560}}\,{\frac{R^{4}}{{t}%
^{4}}}-{\frac{33}{80}}\,{\frac{R^{6}}{{t}^{5}}}-{\frac{67}{60}}\,{\frac{R^{2}%
}{{t}^{3}}})P_{2}(\cos \Theta )  \notag \\
&&\left. +\frac{3}{40}\,\frac{R^{3}}{t^{3}}\,P_{3}(\cos \Theta )+\frac{9}{%
1400}\,\frac{R^{4}}{t^{4}}\,P_{4}(\cos \Theta )\right] ,  \label{GRT}
\end{eqnarray}%
where $t=N/2$, $R=r/2$, and $\Theta $ is the angle between $\mathbf{R}$ and $%
\mathbf{t_{0}}$. The latter is in accordance with the result by Gobush et.
al. \cite{gobush72} derived in a different way. The expansion of $G(R,\Theta
,t)$ for large $N$ can be extended in a straightforward way to include
higher-order corrections.

\section{Numerical Results}

\label{numer}

The computation of the distribution function of the polymer with the fixed
orientation of one end is performed by truncating the infinite order
matrices in (\ref{Gtkp}) with the finite ones, and by taking into account
the finite number of terms in the summation over $l$. The inverse Laplace
transform of (\ref{Gtkp}) is carried out with Maple. The results of the
calculation of the distribution function $G(\mathbf{t}_{0},x=0,y,N)$,
\begin{figure}[tbph]
\includegraphics[scale=0.9]{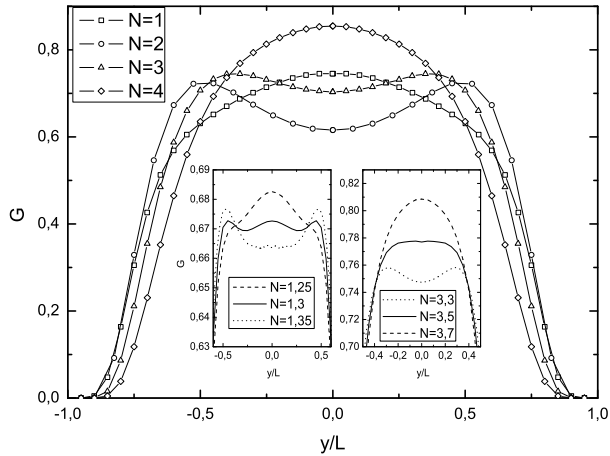}
\caption{Normalized distribution function
$G(\mathbf{t_{0}},x=0,y,N)$ for various chain lengths, computed
with $20\times 20$ matrices. The insets show the distribution
function at the onset of bimodality, and in the region of its
disappearance. } \label{fig-DoublePeak}
\end{figure}
using the truncations with $20\times 20$ order matrices and restricting the
summation over the quantum number $l$ at $l=8$, are given in Fig. \ref%
{fig-DoublePeak}. The results show that the distribution function possesses
the bimodal shape at intermediate chain lengths within the interval $N\in
1.3,...,3.5$. We also find that the distribution function becomes Gaussian
for very short and very long chains. At the onset there is a tiny maximum at
$y=0$, which we interpret as remnant of the Gaussian behavior of short
chains. The maximum at $y=0$ for 3d chain is a rather small effect, which is
difficult to be explained in a qualitative way.

We now will discuss qualitatively the origin of the bimodal behavior of the
projected distribution function of the free end of the wormlike chain. The
very short wormlike chain behaves similar to a weakly bending stiff rod, so
that the distribution function of the free end is Gaussian with the maximum
at $y=0$. The typical conformation of the chain in this regime looks like a
bending rod with constant sign of curvature along the chain. For larger
contour lengths the curvature fluctuations are small and are still
controlled by the bending energy, however with varying sign of curvatures
along the chain. The typical conformation of the chain can be imagined as
undulations along the average conformation of the polymer. The projected
distribution function of the free end in this regime is expected to be
roughly uniform within some range of $y$. We expect that the inhomogeneities
of curvature fluctuations in the vicinity of the clamped end are the reason
for the maximum at $y\neq 0$. The larger curvatures in the vicinity of the
fixed end result in larger displacements $y$ of the free end, and therefore
contribute preferentially to the maximum at $y\neq 0$. With further increase
of $N$ the conformations correspond to undulations around the average
conformation of the chain, which is now a meandering line. Fluctuations
become now less controlled by bending energy, which results in weakening of
the bimodality. Since the difference between 2d and 3d chain is assumed to
be marginal for short chains on the projected distribution function, we will
compare the onset of the bimodality in both cases. Because the transversal
displacement is measured in both cases in units of the contour length, we
have to recompute the number of segments for 3d and 2d chain at the onset
according to
\begin{equation*}
N(3)=\frac{L}{L_{p}(3)}=\frac{L_{p}(2)}{L_{p}(3)}\frac{L}{L_{p}(2)}=\frac{%
L_{p}(2)}{L_{p}(3)}N(2).
\end{equation*}%
Using the dependence of the persistence length on dimensionality \cite%
{benetatos05}, $L_{p}(3)\sim 1/(d-1)$, we obtain $N(3)=2N(2)$. According to
\cite{lattanzi04} $N(2)\approx 0.75$ at the onset, hence we obtain $%
N(3)\approx 1.5$, which is not far from our numerical result, $N(3)\approx
1.3$ (see Fig. \ref{fig-DoublePeak}).

\begin{figure}[tbph]
\includegraphics[scale=0.8]{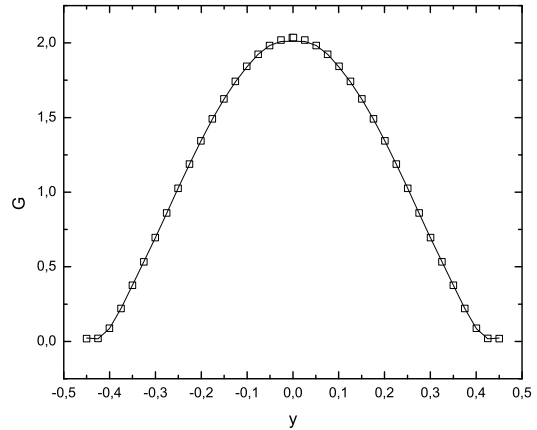}
\caption{Normalized distribution function $G(\mathbf{t_{0}},x=0,y,N)$ for
chain length $N=0.5$. Squares: truncation with $13\times 13$ matrices;
continuous line: truncation with $20\times 20$ matrices. }
\label{fig-N0-5}
\end{figure}

We now will address the issues of accuracy of the calculations, which
depends on the size of the matrices $D$ and $M^{s}$, and the maximal $l$ at
which the summation over $l$ is stopped. In order to check our computation
we verified that at large $N$ ($N=100$) the numerical evaluation of (\ref%
{GyN}) gives with very high accuracy the Gaussian distribution $3/(2N)\,\exp
(-3y^{2}/4N)$. The general tendency is such that the sufficient level of
matrix truncations and the number of terms in the expansion over the
Legendre polynomials increase with decreasing $N$. In the limit $%
N\rightarrow 0$ the whole series over $l$ should be taken into account. The
accuracy of the computations is demonstrated in Fig. \ref{fig-N0-5} showing
the computation of the distribution function for $N=0.5$. The squares and
the continuous curve correspond to the truncations by $13\times 13$ matrices
and $20\times 20$ matrices, respectively. In both cases the summation was
stopped at $l=8$. The corrections due to higher $l$-s are negligibly small.
For example, the corrections associated with $l=10$ and $l=12$ contribute
only in 3rd and 5th decimal digits, respectively. Thus, the computations
depicted in Figs. \ref{fig-DoublePeak}, \ref{fig-N0-5} can be considered as
exact.

\begin{figure}[tbph]
\includegraphics[scale=0.8]{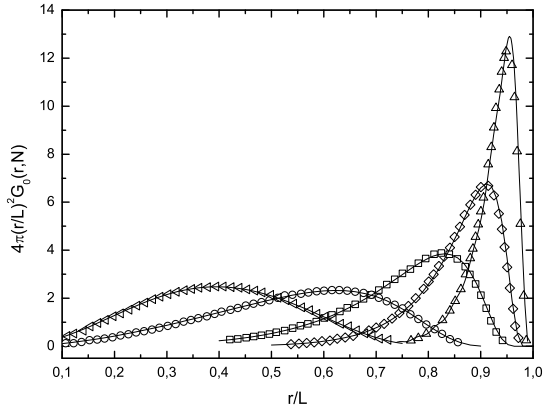}
\caption{The end-to-end distribution function of a free polymer. Solid
lines: the distribution function of a free three dimensional polymer chain
for $N=10$, $5$, $2$, $1$, $0.5$ (from left to right) computed with $%
20\times 20$ matrices. The symbols are the Monte Carlo simulation data
extracted from Fig. 1 in \protect\cite{wilhelm96}.}
\label{figure3}
\end{figure}

Figure \ref{figure3} shows the results of the computation of the 3d
distribution functions $G_{0}(r,N)$ of the free polymer obtained by
performing the inverse Laplace-Fourier transform of the term $l=0$ in Eq. (%
\ref{Gtkp}) for different chain lengths, and its comparison with the Monte
Carlo simulations \cite{wilhelm96}.
Our results are in excellent agreement with the numerical data.

\section{Conclusion}

\label{concl}

To conclude, we have studied the transverse distribution function of the
free end of the three dimensional wormlike chain with fixed orientation and
position of the second end using the exact solution for the Green's function
of the wormlike chain. Within the procedure of truncations of the exact
formula with finite order matrices we find that the distribution function $G(%
\mathbf{t}_{0},x=0,y,N)$ for intermediate chain lengths, belonging to the
interval $1.3L_{p},...,3.5L_{p}$, possesses the bimodal shape with the
maxima at a finite value of the transverse displacement, which is consistent
with the recent studies \cite{lattanzi04,benetatos05} and \cite{spakowitz05}
for the two dimensional chain. In contrast to the 2d wormlike chain, the
transverse 1d distribution function of the 3d chain shows only a tiny peak
at $y=0$ in the vicinity of the onset of bimodality, which however
disappears for larger $N$. We present also results of analytical
considerations for short and large polymers which are in complete agreement
with the classical works \cite{daniels52,gobush72,yamakawa73} where these
limits were investigated using different methods. The computation of the
three dimensional distribution function of a free polymer is in excellent
agreement with the Monte Carlo simulations \cite{wilhelm96}.

\begin{acknowledgments}
A financial support from the Deutsche Forschungsgemeinschaft, SFB 418 is
gratefully acknowledged.
\end{acknowledgments}

\end{document}